# Revisiting Connotations of Digital Humanists: Exploratory Interviews

Ma, Rongqian          Indiana University Bloomington, USA | rm56@iu.edu

**ABSTRACT**

This ongoing study revisits the connotations of "digital humanists" and explores the reasons why a researcher does or does not self-identify as a digital humanist. Building on semi-structured interview data collected from fourteen researchers and practitioners engaging in digital humanities (DH) projects, this poster illustrates researchers' various understandings of "digital humanist" as a term and research identity and highlights the complexity of "digital humanists" as a research community. This study contributes to DH scholarship with insights into the collective imaginations of the "digital humanist" as a research community one decade after the early attempts. Findings of this research study also facilitate a more thorough, timely, and dynamic discussion of the major workforce in digital humanities, potentially paving the way for future research on labor and collaboration in the DH research domain.

**INTRODUCTION AND LITERATURE REVIEW**

This poster revisits a classic question in the digital humanities domain: What are the connotations of "digital humanists?" Early research literature has proposed two major definitions. Alvarado (2012) defined a digital humanist as someone who (1) aims to develop the deep domain knowledge of the traditional humanist, (2) learns a wide variety of technologies and programming languages, and (3) critically situates the technologies as cultural artifacts "participating in the production of social and cognitive structures." This definition of the "digital humanist" requires a scholar to be proficient in both technical skills and humanities knowledge. By contrast, (Ramsay, 2011) argued that a scholar can be called a digital humanist as long as they can build something with digital methods (e.g., applying existing tools or modifying existing codes). This definition embraced a relatively broader reading of the "digital," emphasizing the gradual transition of a humanities scholar into a digital humanist. However, early definitions still consider a digital humanist as primarily a *humanist* with strong humanities training. Such definitions captured different perceptions of "digital humanists" as a research community in the early stage.

The recent technological advancement has expanded the reach of digital humanist as a research community and raised new questions about the connotations of the term. *How do DH researchers think of the role of technological literacy in their work and research identify? Is a digital humanist still primarily a humanist researcher, or does a digital humanist need formal humanities training?* Jänicke (2016) called for a broader, more inclusive conception of the digital humanists community, arguing that the DH field would benefit from the active participation of researchers from various knowledge domains, especially those from scientific fields. Recent research in the landscape of DH illustrated that "DH is simultaneously a discipline in its own right and a highly interdisciplinary field, with many connecting factors to neighboring disciplines—first and foremost, computational linguistics, and information science" (Luhmann & Burghardt, 2022). In addition to the roles of DH among academic fields, Ma and Li (2022) also demonstrated that humanities and STEM researchers, among others, further shape the DH community with their respective, sometimes competing, research conventions as well as cross-field research collaborations. Empirical works focusing on DH intellectual structures, disciplinary compositions, and collaboration also suggest an increasingly complex picture of DH workforce (Antonijević, 2015; Fiormonte et al., 2015; Griffin & Hayler, 2018; Papadopoulos & Reilly, 2020).

The increasing "complexity" of DH workforce has raised the urgency to revisit the classic question in DH: "Who is in and who is out?" (Ramsay, 2011). Using semi-structured interview method, this study explores the current understandings and conceptions of a "digital humanist," illustrating why and why not a researcher self-identifies as a digital humanist. Scholarly literature has also shown that active DH researchers do not always think of themselves as "digital humanists" (Burdick, 2012). Rich qualitative data collected from the interviews will help highlight the major perceptions of what a digital humanist means. Ten years into the discussion of the research identities of digital humanists, the findings of this exploratory study will also inform future research on DH workforce, teams, and cross-field collaboration.

**DATA AND METHODS**

**Participant**

Fourteen participants were identified through a snowball sampling technique based on the author's personal network (Biernacki & Waldorf, 1981). Any researcher or practitioner who has been actively engaging in DH projects can be qualified as interview candidates. Using this inclusive screening criterium, this study recruited fourteen participants representing a diverse pool in terms of disciplines, academic levels, and seniority. The represented fields and disciplines include, for example, history, religious studies, English, art history, sociology, information science, math, and anthropology. The participants also represent various levels of training and research experience, including master's students (n=2), PhD candidates (n=5), postdoc (n=1), assistant professors (n=4), associate professors (n=1), and curator (n=1). Among the fourteen interviewees, eight participants (P2, P4, P6, P7, P9, P10, P11, P14) self-identified as digital humanists while six (P1, P3, P5, P8, P12, P13) did not.

### Interview Design and Analysis

Semi-structured interview method was used for this study. Semi-structured interviews have the unique advantage of preserving participants' rich descriptions and detailed responses, especially the original discussions of "digital humanists" and the reasons for their self-identification as digital humanists. Each interview lasted for 45 minutes and consisted of two sections. In the first section, I collected structured demographic information from the participants, such as the participant's current academic position, major field(s) of study, and if they would self-identify as a digital humanist. The second section discussed the participant's self-identification, focusing on the reasons for their choices and how they understand the meaning of "digital humanists." The interviews were audio-recorded and fully transcribed for analysis. Qualitative content analysis was then used to code and analyze the participants' responses, particularly their accounts on the identities of digital humanists.

## RESULTS

Preliminary results of this study captured various understandings of "digital humanists" among researchers. **Technological skills define a digital humanist.** Results show that some researchers (P2, P6, P9) evaluate their eligibility of being a "digital humanist" based on their technical skills, especially the ability to write codes or run computational models. They claimed that it is the programming ability that made them a "digital humanist" rather than a "humanist." Some participants, however, argued that **actual work experiences define a digital humanist**. P4, for example, said "I self-identify as a digital humanist in the same way as I self-identify as an information designer, a programmer, and a statistician. I do research and work in digital humanities, so I guess *technically* I can say I am a digital humanist." This account treats "digital humanist" as a non-mutually exclusive, fluctuating status that reflects current work state and content. P10 also considered herself as a digital humanist because she had supervised students in DH projects. Compared with the former group who understand digital humanists based on training and skillsets, scholars and practitioners in the latter group believed one can be a digital humanist "by doing it," emphasizing the actual practices over the conceptual, artificial definitions of the term.

The preliminary results also suggest that some researchers did *not* self-identity as digital humanists because they are concerned with the **nature of their research questions and the necessity of using digital methods in research**. **For instance,** P1 and P5 did *not* self-identify as digital humanists because they consider their research questions "not ultimately bond with digital methods." A museum art curator trained in art history, museum studies, and sociology, P3 was concerned that new constructs such as "DH" did not generate fundamentally new questions distinct from those in conventional or analog humanities, which made him reluctant to self-identify as a "digital humanist," rather than just a humanist. From a different perspective, P12 and P13 did not self-identify as digital humanists because they would not think of themselves as "humanists," who, according to their accounts, must have humanities training and work on humanities research questions in an in-depth manner. Both trained and working as information scientists, P12 and P13 demonstrated that they have been engaging in DH projects from a technological perspective, either by means of working as a programmer (P12) in the team or by solving problems related to digital methodologies (e.g., how to retrieve and search music information, or create digital simulations for cultural heritage sites, P13).

## CONCLUDING REMARKS

The preliminary findings of this study suggest that while technological literacy is still an important factor that impacts researchers' self-identification as a digital humanist, researchers have also developed a more inclusive understanding of the term, which focuses more on the actual DH work engagement rather than the "qualifications" (e.g., if they can code or if they have humanities degrees). These researchers acknowledge that DH work can be of different shapes and each researcher can contribute in a unique and meaningful way – and hence, call themselves as digital humanists. Such a broader view of digital humanists as a research community has potential benefits. It may further facilitate researchers to embrace research methods and paradigms outside of their home fields, as well as to welcome new collaborators, who, for example, speak different disciplinary languages or have distinct ways of thinking. Despite its values, this study bears limitations. Fourteen interviews are not sufficient to conclude all considerations that go into the current definitions of, and researchers' self-identification as, digital humanists. In future work, to address the limitation, more participants will be recruited and the survey method will be applied to expand the sample size. Analysis of a larger dataset with richer details will help develop a more thorough idea of the current conceptions of "digital humanists" among researchers engaging in DH projects.

**ACKNOWLEDGMENT**
The author would like to thank all the interview participants for their generosity in sharing thoughts, experiences, and ideas. This study would not be possible without their contribution.**REFERENCES**

Alvarado, R. C. (2012). The digital humanities situation. In M. Gold (Ed.), *Debates in the Digital Humanities*. University of Minnesota Press.

Antonijević, S. (2015). *Amongst digital humanists: An ethnographic study of digital knowledge production*. Palgrave Macmillan.

Biernacki, P., & Waldorf, D. (1981). Snowball sampling: Problems and techniques of chain referral sampling. *Sociological Methods & Research*, *10*(2), 141–163.

Burdick, A. (Ed.). (2012). *Digital humanities*. MIT Press.

Fiormonte, D., Numerico, T., & Tomasi, F. (2015). *The Digital Humanist: A Critical inquiry*. Punctum Books. https://doi.org/10.21983/P3.0120.1.00

Griffin, G., & Hayler, M. S. (2018). Collaboration in Digital Humanities Research – Persisting Silences. *Digital Humanities Quarterly*, *012*(1).

Jänicke, S. (2016). Valuable Research for Visualization and Digital Humanities: A Balancing Act. *Workshop on Visualization for the Digital Humanities, IEEE VIS 2016, Baltimore, Maryland, USA*.

Luhmann, J., & Burghardt, M. (2022). Digital humanities—A discipline in its own right? An analysis of the role and position of digital humanities in the academic landscape. *Journal of the Association for Information Science and Technology*, *73*(2), 148–171. https://doi.org/10.1002/asi.24533

Ma, R., & Li, K. (2022). Digital humanities as a cross-disciplinary battleground: An examination of inscriptions in journal publications. *Journal of the Association for Information Science and Technology*, *73*(2), 172–187. https://doi.org/10.1002/asi.24534

Papadopoulos, C., & Reilly, P. (2020). The digital humanist: Contested status within contesting futures. *Digital Scholarship in the Humanities*, *35*(1), 127–145. https://doi.org/10.1093/llc/fqy080

Ramsay, S. (2011). Who's in and who's out. In M. Terras, J. Nyhan, & E. Vanhoutte (Eds.), *Defining digital humanities: A reader*. Ashgate.ASIS&T Annual Meeting 2022    3    Submission Type